\documentclass[aps,prl,groupedaddress,twocolumn]{revtex4-1}
\usepackage{bm}
\usepackage{amssymb}
\usepackage{amsmath}
\usepackage{graphicx}
\usepackage{braket}
\newcommand{\SUd}{{{SU($d$)}}}

\newcommand{\bmt}{{\ensuremath{\bm{\theta}}}}
\newcommand{\bmm}{{\ensuremath{\bm{\mu}}}}

\newcommand{\bmx}{{\ensuremath{\bm{\chi}}}}
\newcommand{\bmg}{{\ensuremath{\bm{g}}}}
\newcommand{\bmI}{{\ensuremath{\bm{I}}}}

\newcommand{\RN}{{\ensuremath{\mathbb{R}^N}}}
\newcommand{\allones}{{\ensuremath{\vec{\mathbf{1}}}}}

\begin{document}

\title{Randomized Benchmarking, Correlated Noise, and Ising Models}

\author{Bryan H. Fong}
\email[]{bhfong@hrl.com}
\author{Seth T. Merkel}
\affiliation{HRL Laboratories, LLC, 3011 Malibu Canyon Road, Malibu, CA 90265}

\date{\today}

\begin{abstract}
We compute the expected randomized benchmarking sequence fidelity for a system
subject to Gaussian time-correlated noise. For single qubit benchmarking we show that
the expected sequence fidelity is given by the partition function of a
long-range coupled spin-one Ising model, with each site in the Ising model
corresponding to a free evolution interval.  For
$d$-state systems, the expected sequence fidelity is given by an
Ising-like model partition function whose site variables
are given by the weights of the adjoint representation of \SUd.
A high effective temperature
expansion for the partition function in the single qubit case
shows decay of sequence fidelity varying
from exponential for uncorrelated noise to a power law for quasistatic noise.
Fitting an exponential to the sequence fidelity decay under correlated noise
gives unreliable estimates of the average gate error rate.  \end{abstract}

\maketitle

Randomized benchmarking (RB)
\cite{Knill:2008fi,Magesan:2011bx,Magesan:2012to,Magesan:2012tn,Epstein:2014oe}
has become a standard method for characterizing
gate error rates in quantum computing. The RB
protocol is simple:
prepare an initial state,
apply a sequence of quantum gates, measure the fidelity of the final state, repeat
with sequences of increasing length, and fit the fidelity versus sequence
length to an exponential to obtain the average gate error rate.
The simplicity and efficiency of the protocol has led to its
widespread use in many qubit technologies, including
superconductors \cite{Chow:2009kv,Magesan:2012to,Barends:2014bq},
ions \cite{Knill:2008fi,Gaebler:2012kt,Harty:2014df},
solid state quantum dots \cite{Veldhorst:2014pd,Fogarty:2015vo}, and
atomic nuclei \cite{Pla:2013az}.
Though the protocol was originally developed to characterize uncorrelated
Markovian
noise, the qubit systems where RB has been used are
generally subject to non-Markovian correlated noise
\cite{Wellstood:1987qt,Yoshihara:2006wk,
Bylander:2011kl,Slichter:2012am,Anton:2013hm,
Taylor:2007im,Veldhorst:2014pd,Eng:2015kr,Fogarty:2015vo}.
In this paper we analyze the effects of correlated noise on RB.
We show a surprising
formal equivalence between single qubit RB and a long-range coupled spin-one
Ising model.
The connection to the Ising model leads to a determinant formula
that gives RB sequence fidelity decays ranging from
exponential to power law, depending on noise correlations.

The effects of correlated noise on RB have previously
been examined, both analytically and numerically.
Magesan et al.\ \cite{Magesan:2012tn}
as well as Wallman and Flammia \cite{Wallman:2014xc} demonstrated
the robustness of the RB estimated error rate
to weakly time dependent gate noise. For $1/f$ correlated noise,
Epstein et al.\ \cite{Epstein:2014oe} showed through numerical simulations
that RB gives an error rate within
a factor of two of the true average gate error rate.
For general Hamiltonian-driven correlated noise,
Ball et al.\ \cite{Ball:2016cs} derived sequence fidelity probability density
functions valid to linear order in the product of sequence length $N$ and
average gate error rate $\varepsilon$.
The determinant formula for the average sequence fidelity that we derive
is valid for general $N$, allowing us to quantify the degree to which sequence
fidelity is nonexponential.

Our analysis shows that the expected RB sequence fidelity
is given by a spin-one Ising model partition function.
The Ising model has effective coupling strengths given by the
covariance matrix of error phases accumulated in benchmarking intervals
and an effective temperature given by the inverse of the gate error rate.
We use the techniques of statistical
field theory \cite{Amit:2005lk} to obtain a high temperature series expansion
for the RB sequence fidelity,
and also to make explicit the relationship between
RB and random dynamical decoupling
\cite{Viola:1999fe,Viola:2005rp}.
The lowest order term in the high temperature expansion
takes the form of a finite rank Toeplitz determinant \cite{Fisher:2007zp},
giving $N$ dependence in the sequence fidelity varying
from exponential for uncorrelated noise to power law for quasistatic noise.
Because fitting the power law decay to an exponential
produces unreliable results, we propose an alternative
fitting procedure based on the observation that the initial fidelity decay
is independent of noise correlations \cite{Ball:2016cs}.
Finally, we show that the formal equivalence between RB and
long-range coupled Ising models extends to benchmarking of $d$-state systems, with
Ising model site variables that are the $d^2 - 1$ weights of the adjoint
representation of \SUd.

The expected value of the RB sequence fidelity $P_0$
is given by the noise-averaged $N$-fold composition of the twirled
(group-averaged) free evolution
operator $\mathcal{R}$:
\begin{equation}
  P_0 = \left\langle \mathrm{tr}\left(\rho_0.
  \mathcal{R}^{(N)}\circ\ldots\circ\mathcal{R}^{(2)}\circ\mathcal{R}^{(1)}(\rho_0)\right)
   \right\rangle_\mathrm{noise},
\label{eq:P0}
\end{equation}
for an initial density matrix $\rho_0$ that is a pure state. The twirled
free evolution operator in interval $n$ is defined as
\begin{equation}
\mathcal{R}^{(n)}(\rho) = \int_{\mathrm{U}(d)} dU U^\dagger F_n U
\rho U^\dagger F_n^\dagger U,
\label{eq:RDefinition}
\end{equation}
where U($d$) is the $d$-dimensional unitary group, $dU$ is the Haar measure
for U($d$), and $F_n$ is the noisy unitary free evolution in interval $n$. The
twirled free evolution operator is implemented experimentally using a unitary 2-design
for U($d$), converting the integral over the Haar measure in Eq.~(\ref{eq:RDefinition})
to a finite sum over the design. For qubit systems, the 2-design is usually the
Clifford group. Eq.~(\ref{eq:P0}) then contains two averages: a noise average
and a group (or sequence) average. We assume that these averages are independent
so that each sequence sees all noise realizations. Experimentally, there may be
noise correlations between sequence executions not fully captured by the present
formalism.

For a single qubit subject to Hamiltonian noise, the unitary free evolution operator
is $F_n = e^{-i \frac{\theta_n}{2} \hat{m}.\boldsymbol{\sigma}}$,
where $\theta_n$ is the error phase accumulated in free evolution interval
$n$, $\hat{m}$ is a unit vector giving the axis of rotation in 3-space,
and $\boldsymbol{\sigma}$ is the vector of Pauli matrices. Taking as a basis for the
space of density matrices the identity matrix $\sigma_0$ and Pauli matrices
$\{\sigma_1,\sigma_2,\sigma_3\}$, the matrix representation of the
twirled free evolution operator Eq.~(\ref{eq:RDefinition}) is
$R^{(n)}_{00}=1$, $R^{(n)}_{ij}=\frac{1}{3}(1+2\cos\theta_n)\delta_{ij}$ for
$i,j = 1,2, \mathrm{or\ } 3$; all
other components are 0 \cite{Epstein:2014oe}.
For any initial pure-state density matrix
the repeated
application of the twirled free evolution map results in a sequence fidelity of
\begin{equation}
  P_0 = \frac{1}{2} + \frac{1}{2}
  \left\langle\prod_{n=1}^{N}\frac{1}{3}\left(1+2\cos\theta_n\right)
  \right\rangle_{\mathrm{noise}}
  \equiv \frac{1}{2} + \frac{1}{2}\mathcal{Z},
\label{eq:singleQubitP0}
\end{equation}
where we have defined $\mathcal{Z}$ as the noise-averaged product.

Assuming that the length $N$ vector of accumulated error phases $\bmt=
\{\theta_n\}_{n=1}^N$ is Gaussian distributed,
\begin{equation}
  \mathcal{Z} = \int_{\RN} d\bmt
  \left[
  \prod_{n=1}^{N}\frac{1}{3}\left(1+2\cos\theta_n\right)
  \right]
  \frac{e^{-\frac{1}{2}(\bmt-\bmm).\bmx^{-1}.(\bmt-\bmm)}}{\sqrt{(2\pi)^N |\bmx|}},
\label{eq:singleQubitZ}
\end{equation}
where $\bmm$ is the vector of mean
accumulated error phases, $\bmx$ is the positive semi-definite
covariance matrix of the noise, and $|\bmx|$ is the determinant of $\bmx$.
In the case that $\bmx$ has zero eigenvalues,
we take the distribution as giving Dirac delta functions in the directions of
the eigenvectors whose associated eigenvalues are zero.
We assume that the mean accumulated error
phase in each interval is the same, so that $\bmm = \theta_0 \allones$, where
$\allones$ is the length $N$ vector of all ones.
Before evaluating $\mathcal{Z}$ for general
covariance matrices, we examine the uncorrelated (Markovian)
and quasistatic noise limits.

For noise that is uncorrelated between intervals and identical on each interval,
the covariance matrix is $\bmx = 2\beta \bmI$, where $\beta$ parameterizes the
strength of the noise ($\beta=3\varepsilon$, with $\varepsilon$
the average gate error rate) and $\bmI$ is the $N\times N$ identity matrix.
Because the covariance matrix is diagonal, the $N$-dimensional
integral in Eq.~(\ref{eq:singleQubitZ}) becomes the product of $N$ identical
integrals:
\begin{equation}
\mathcal{Z}_\mathrm{uncorrelated} =
\left(
\frac{1+2 e^{-\beta}\cos\theta_0}{3}
\right)^N,
\label{eq:exactUncorrelatedZ}
\end{equation}
leading to the standard exponential decay in sequence fidelity for
DC and/or uncorrelated noise. For quasistatic noise, the covariance matrix
$\bmx = 2\beta \allones\otimes\allones$ is singular: it has one eigenvalue
of $2 \beta N$ and $N-1$ zero eigenvalues.
Integrating over the delta functions
associated with the zero eigenvalues leaves the single integral:
\begin{eqnarray}
\mathcal{Z}_\mathrm{quasistatic} &=&
  \int_\mathbb{R} d\theta \left[
  \frac{1+2\cos\theta}{3}  \right]^N
  \frac{e^{-\frac{(\theta-\theta_0)^2}{4 \beta}}}{\sqrt{2\pi2 \beta}}\\
&&\hspace*{-2cm}=\frac{1}{3^N}\left[
\binom{N}{0}_2 + 2 \sum_{k=1}^N \binom{N}{k}_2 e^{-\beta k^2}\cos k\theta_0
\right],
\label{eq:exactQuasistaticZ}
\end{eqnarray}
where $\binom{N}{k}_2$ is a trinomial coefficient \cite{Euler:1911hs}.
These exact expressions for uncorrelated and quasistatic noise are useful
in assessing the accuracy of approximate expressions for the sequence
fidelity presented below.
For both uncorrelated and quasistatic noise,
the case of a single free evolution interval $N=1$
allows us to identify $\beta$ as parameterizing the envelope decay of a driven
oscillation $\cos\theta_0$.
Defining $\tau$ to be the length in time of the free
evolution interval, $\beta$ is related to $T_2^*$ through
$\beta \approx (\frac{\tau}{T_2^*})^\alpha$, with the exponent $\alpha$
and characteristic noise time $T_2^*$ depending
on the noise process, while $\theta_0 = \omega \tau$ for a DC field characterized
by strength $\omega$.

Returning to the general expression Eq.~(\ref{eq:singleQubitZ}), we rewrite
the product of $1+2\cos\theta_n$ terms as a sum of cosines:
\begin{equation}
  \prod_{n=1}^N 1+2\cos\theta_n = \sum_{\bmg \in \{-1,0,1\}^N} \cos \bmg.\bmt.
\end{equation}
Here $\bmg$ is a length $N$ vector each of whose components is $1$, $-1$, or $0$;
the sum is over all $3^N$ configurations of $\bmg$. The validity of this equality
can be shown by induction using the cosine addition formula.
Eq.~(\ref{eq:singleQubitZ}) then takes the form
\begin{eqnarray}
  \mathcal{Z} &=& \frac{1}{3^N}\sum_{\bmg \in \{-1,0,1\}^N} \int_\RN d\bmt
  \cos\bmg.\bmt
  \frac{e^{-\frac{1}{2}(\bmt-\bmm).\bmx^{-1}.(\bmt-\bmm)}}{\sqrt{(2\pi)^N |\bmx|}}
\label{eq:HSTransformation}
  \\
  &=&  \frac{1}{3^N}\sum_{\bmg \in \{-1,0,1\}^N}
  e^{-\frac{1}{2}\bmg.\bmx.\bmg}e^{i \bmg.\bmm}.
  \label{eq:singleQubitPartitionFunctionSum}
\end{eqnarray}
The equality of Eqs.~(\ref{eq:HSTransformation}) and (\ref{eq:singleQubitPartitionFunctionSum})
is the Hubbard-Stratonovich transformation
\cite{Stratonovich:1957uq,Hubbard:1959fk}, used to convert an Ising model
to an associated field theory. We use the inverse transformation here,
converting the ``field theory'' of the sequence fidelity to an Ising model.
The right hand side of Eq.~(\ref{eq:singleQubitPartitionFunctionSum}) is
a partition function for an $N$-site spin-one Ising model with long-range
coupling, normalized to an infinite temperature value of $3^N$.
Interactions between site variables have coupling strength
determined by the covariance matrix $\bmx$, and the $n^\mathrm{th}$ site couples
to an imaginary magnetic field $\mu_n$. If both $\bmx$ and $\bmm$
scale with $\beta$, we can identify $\beta$ as the inverse temperature.
Because it characterizes the noise strength,
$\beta \approx (\frac{\tau}{T_2^*})^\alpha$ must be
small, corresponding to the high temperature limit;
otherwise a RB experiment would not be useful.
Though
Eq.~(\ref{eq:singleQubitPartitionFunctionSum}) has the form of a partition
function sum, it is \emph{not} used in the same way that a normal
partition function is used. The expected RB sequence
fidelity is directly proportional to $\mathcal{Z}$: $\mathcal{Z}$ is not used
as a probability density function whose normalization is immaterial,
but instead the actual value of $\mathcal{Z}$ is its key property.

Equation~(\ref{eq:singleQubitPartitionFunctionSum}) reveals a pleasing connection
between single qubit RB and a spin-one Ising model
partition function. However, using Eq.~(\ref{eq:singleQubitPartitionFunctionSum})
to determine the sequence fidelity for
given $\bmx$ and $\bmm$ requires evaluating a sum with a number of terms
exponential in the sequence length $N$.
Using $\beta$ as an expansion parameter, we perform
a high temperature/weak coupling expansion of Eq.~(\ref{eq:singleQubitZ})
to derive an approximate
expression for the single qubit sequence fidelity, presented
in the supplementary material.
When the mean noise $\theta_0=0$,
the approximate form for the partition function is
\begin{equation}
  \mathcal{Z}_0 = \frac{1}{\sqrt{\left|\bmI + \frac{2}{3}\bmx\right|}}
  \left(1-\frac{1}{12}\sum_{n=1}^N\Sigma_{nn}^2+\ldots\right),
\label{eq:singleQubitZ0Approximate}
\end{equation}
where $\Sigma\equiv\bmx(\bmI+\frac{2}{3}\bmx)^{-1}$.
The leading order dependence of the sequence fidelity depends only
on the determinant of a symmetric positive definite matrix;
for wide-sense stationary noise this matrix is also Toeplitz.
Since the determinant expression
is common to all terms in the $\mathcal{Z}$ series expansion, the correction terms
within the parentheses specify relative errors compared to the leading order term.
The first correction term nominally has $N\beta^2$ scaling, with
the actual $N$ scaling dependent on the covariance matrix $\bmx$.

For uncorrelated and quasistatic
noise the Toeplitz determinants are readily computed analytically.
For the case of uncorrelated noise $\bmx=2\beta\bmI$,
\begin{equation}
  \mathcal{Z}_{0,\mathrm{uncorrelated}} = \left(1+\frac{4}{3}\beta\right)^{-\frac{N}{2}}
  \left(1-\frac{N\beta^2}{3(1+\frac{4}{3}\beta)^2}
  +\ldots
  \right),
  \label{eq:approximateUncorrelatedZ}
\end{equation}
and the sequence fidelity decays exponentially with increasing $N$, with error
rate the same to $\mathcal{O}(\beta)$ as that given by
Eq.~(\ref{eq:exactUncorrelatedZ}).
The previously stated $N\beta^2$ dependence of the correction term is
apparent. Though the relative error may become
large for large $N\sim\beta^{-2}$, the absolute error will be exponentially
small because of the prefactor. For the case of quasistatic noise,
$\bmx=2\beta \allones\otimes\allones$, and
\begin{equation}
\mathcal{Z}_{0,\mathrm{quasistatic}} = \frac{1}{\sqrt{1+\frac{4}{3}N\beta}}
\left(1-\frac{N\beta^2}{3\left(1+\frac{4N\beta}{3}\right)^2}
\ldots\right).
\label{eq:approximateQuasistaticZ}
\end{equation}
The quasistatic noise sequence fidelity decays as the inverse square root of
$N$, significantly slower than in the case of uncorrelated noise.
The correction term now scales as $1/N$ for large $N$, so that both
the relative and absolute errors of the leading order quasistatic series expansion
compared to the exact expression Eq.~(\ref{eq:exactQuasistaticZ}) decrease
with large $N$.
One may compute
additional terms of the series expansion in Eq.~(\ref{eq:singleQubitZ0Approximate})
if high accuracy in sequence fidelity is required.

\begin{figure}
\includegraphics[width=3in]{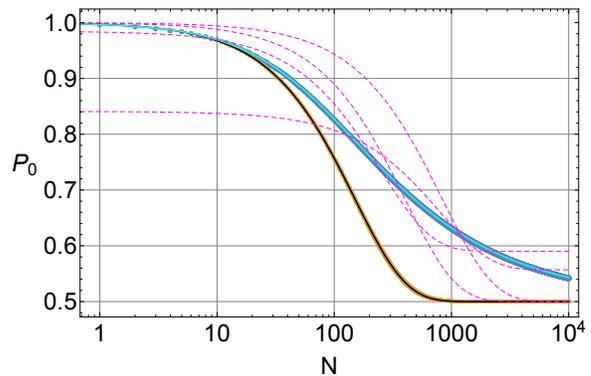}
\caption{\label{fig:decayPlot}
Sequence fidelity $P_0$ as a function of RB sequence
length $N$. The orange data shows uncorrelated noise sequence fidelity
corresponding to Eq.~(\ref{eq:exactUncorrelatedZ}) for $\beta=0.01$ and
$\theta_0=0$; the black curve is best fit to the orange data using $A+B(1-2\varepsilon)^N$.
The dark blue data shows quasistatic noise sequence fidelity corresponding to
Eq.~(\ref{eq:exactQuasistaticZ}) for $\beta=0.01$ and $\theta_0=0$;
the cyan curve is best fit to
the dark blue data using the leading order term in Eq.~(\ref{eq:approximateQuasistaticZ}).
Magenta curves
are best fits to the quasistatic dark blue data using $A+B(1-2\varepsilon)^N$ for four
different fitting scenarios, discussed in the text.
}
\end{figure}

Uncorrelated and quasistatic noise display significantly different decay
dependence on $N$. Figure \ref{fig:decayPlot}
shows the decay of sequence fidelity for
uncorrelated and quasistatic noise computed from the exact
expressions Eqs.~(\ref{eq:exactUncorrelatedZ})
and (\ref{eq:exactQuasistaticZ}), respectively, for the same noise parameter
$\beta=0.01$ and $\theta_0=0$.
Fitting the uncorrelated noise decay to the
standard RB decay expression $A+B(1-2\varepsilon)^N$
recovers the original noise parameter $\beta=0.01$;
fitting the quasistatic noise decay to
the leading order decay in Eq.~(\ref{eq:approximateQuasistaticZ})
also recovers $\beta=0.01$.
Fitting quasistatic noise decay to
$A+B(1-2\varepsilon)^N$
gives inconsistent estimates of the noise parameter $\beta$.
Magenta curves in Fig.~\ref{fig:decayPlot} show four fitting
scenarios, with $A$ and $B$ held fixed or allowed to vary, and
with equal weighting of
sequence fidelity data
or weighting by inverse sequence length in the least squares objective.
All curves shown here underestimate the average gate error rate by factors
of two to six; overestimation
is also possible if only short RB sequence lengths are used for fitting.
A more detailed analysis of fitting power law decay to an exponential is
given in the supplementary information.
This example demonstrates the danger in
assuming that all benchmarking experiments can be fit to exponential decay.

If the form but not the overall scale $\beta$ of the covariance matrix $\bmx$ is known,
fitting benchmarking data to Eq.~(\ref{eq:singleQubitZ0Approximate}) estimates
$\beta$. With no prior knowledge of $\bmx$,
an alternative procedure for determining $\beta$ or $\varepsilon$ emerges from the
observation that the sequence fidelity for small $N$ is
independent of the noise correlations, a result previously
noted in \cite{Ball:2016cs}.
To lowest order in $N\beta$, Eq.~(\ref{eq:singleQubitZ0Approximate})
gives $\mathcal{Z}_0 \approx 1-\frac{2}{3} N\beta$, independent of all the off-diagonal
elements of the covariance matrix. The initial decay in sequence fidelity can
be fit to $A + B(1-2 N \varepsilon)$, where $A$ is determined by the asymptotic value of
$P_0$ for large $N$. Fitting the quasistatic expression
Eq.~(\ref{eq:exactQuasistaticZ}) shown in Fig.~\ref{fig:decayPlot}
with $A=P_0(N=10^4)=0.543$ yields $\beta = 0.0107$, a 7\% relative error
in the average gate error rate. Errors in fitting $\beta$ of
$\mathcal{O}(\beta^2)$ and $\mathcal{O}(\beta\delta/B)$ arise from the
linear approximation and the error $\delta$ in the asymptotic value for $A$,
respectively.

If we assume that the noise is limited to a single axis, the covariance matrix
$\bmx$
can be expressed simply in terms of the noise power spectral density (PSD).
For dephasing
noise with Hamiltonian $H=\frac{1}{2} \sigma_z B(t)$, the error phase
accumulated in a
free evolution interval is $\theta_n = \int_{(n-1)\tau}^{n\tau}dsB(s)$.
Using the Wiener-Khinchin theorem, the covariance matrix
components for $m$,$n\in 1,\ldots,N$ are
\begin{eqnarray}
  \chi_{mn} &=& \left\langle (\theta_m-\theta_0)(\theta_n-\theta_0)
  \right\rangle_\mathrm{noise}
\nonumber\\
  &=& \int_0^\infty df \frac{\left[\cos\left(2\pi f\tau(m-n)\right)\sin^2\pi f\tau\right]}
  {\pi^2f^2} S(f),
  \label{eq:covariancePSD}
\end{eqnarray}
where $S(f)$ is the PSD of $B(t)$. Define $\phi_{mn}$ to be the quantity in the
square brackets in Eq.~(\ref{eq:covariancePSD}).  With the effective Ising model
Hamiltonian in Eq.~(\ref{eq:singleQubitPartitionFunctionSum}) as $H_\mathrm{Ising}
=\frac{1}{2} g_m\chi_{mn}g_n -i g_n \mu_n$, we recognize
$\frac{1}{2}g_m\phi_{mn}g_n$ as giving filter functions used in dynamical
decoupling analyses
\cite{Ball:2016cs,Cywinski:2008ys,Biercuk:2011zr,Green:2012rb,Green:2013cr}.
For a dynamical decoupling
pulse sequence of length $N$, a single configuration of site variables
$\{g_n\}_{n=1}^N$ taking only values of $1$ or $-1$ specifies the decoupling
sequence and associated filter function $F(f\tau)=\frac{1}{2}g_m\phi_{mn}g_n$.
In the RB context, we can interpret the
partition function sum Eq.~(\ref{eq:singleQubitPartitionFunctionSum})
and corresponding sequence fidelity Eq.~(\ref{eq:singleQubitP0})
as the average fidelity over $3^N$ ``decoupling'' (really randomization)
sequences.
An additional value of $0$ for each site variable is permitted by the
twirled free evolution map, in contrast to dynamical decoupling where only
``forward'' (1) and ``echoed'' ($-1$) intervals are allowed.  This is because
the twirled
free evolution map takes $g_m$ values from the weights of the adjoint
representation of SU(2) while single axis dynamical decoupling takes
$g_m$ values from the non-trivial irreducible representation of $\mathbb{Z}_2$.

\begin{table}
\caption{\label{tab:PSDTable}
PSD parameters for Eq.~(\ref{eq:PSD}) and corresponding free evolution decay
parameters. Colors refer to Fig.~\ref{fig:PSDDecayPlot}.
}
\begin{ruledtabular}
\begin{tabular}{llllll}
& $f_L[\mathrm{Hz}]$ & $f_H[\mathrm{Hz}]$ & $A[\mathrm{Hz}]$ & $T_2^*[\mathrm{s}]$ & $\alpha$\\
orange (uncorrelated) &$\infty$ & -- & $2.0\times 10^6$ & $1.0\times10^{-6}$&1\\
brown    & $10^8$   & $10^{10}$ & $4.2\times10^6$    & $9.5\times10^{-7}$ & 1\\
purple   & $10^6$   & $10^{10}$ & $4.3\times10^7$    & $1.0\times10^{-7}$ & 2\\
red      & $10^{-3}$& $10^{10}$ & $7.9\times10^{15}$ & $1.0\times10^{-7}$ & 2\\
green    & $10^{-3}$& $10^{5}$  & $9.8\times10^{15}$ & $1.0\times10^{-7}$ & 2\\
blue (quasistatic) &-- &--      & --                 & $1.0\times10^{-7}$ & 2 \\
\end{tabular}
\end{ruledtabular}
\end{table}

\begin{figure}
\includegraphics[width=3in]{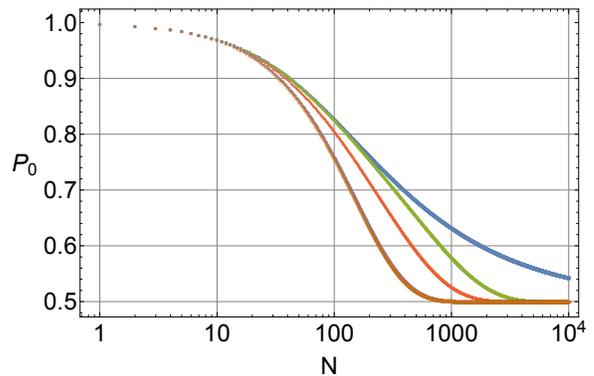}
\caption{\label{fig:PSDDecayPlot}
Sequence fidelity $P_0$ vs. RB sequence
length $N$. All data has $\beta=0.01$, $\theta_0=0$, and $\tau=10^{-8}\mathrm{s}$.
Orange data is for
uncorrelated noise, Eq.~(\ref{eq:exactUncorrelatedZ}), and blue data for
quasistatic noise, Eq.~(\ref{eq:exactQuasistaticZ}).  Brown, purple, red and green
data correspond to PSDs with parameters given in Table~\ref{tab:PSDTable} and
functional form in Eq.~(\ref{eq:PSD}). Brown, purple,
and orange data are nearly indistinguishable on the plot.
}
\end{figure}

For a PSD of the form \cite{Paladino:2014kx}
\begin{equation}
S(f) = \begin{cases}
A       & 0\leq f<f_L\\
A\frac{f_L}{f}   & f_L\leq f \leq f_H\\
A\frac{f_L f_H}{f^2} & f_H <f,
\end{cases}
\label{eq:PSD}
\end{equation}
a variety of different sequence fidelity decay behaviors is possible.
This continuous, piecewise PSD is constant at low frequency, $1/f$ at medium
frequencies, and $1/f^2$ at high frequencies.
For a PSD of this functional form the covariance matrix
integrals Eq.~(\ref{eq:covariancePSD}) have (large) closed
form expressions as
functions of $f_L$, $f_H$, $A$, and $\tau$, with which approximate sequence fidelities
are computed using Eq.~(\ref{eq:singleQubitZ0Approximate}).
Figure~\ref{fig:PSDDecayPlot} shows the sequence fidelity decay for
four choices for the $A$, $f_L$, and $f_H$ parameters given
in Table~\ref{tab:PSDTable}; all sequence fidelities are computed with
$\tau=10^{-8}\mathrm{s}$.
Uncorrelated (orange) and quasistatic (blue) sequence fidelity data\
are shown for comparison
in Fig.~\ref{fig:PSDDecayPlot},
with all six curves having $\beta=0.01$ and $\theta_0=0$. From
$\beta=0.01\approx(\frac{\tau}{T_2^{*}})^\alpha$, one can determine $T_2^{*}$
and $\alpha$ from a $\tau$-series expansion of $\chi_{11}$.
$T_2^{*}$
and $\alpha$ for the different PSDs are also given in Table~\ref{tab:PSDTable}.
Since all six curves
have the same $\beta$ value, small $N$ behavior is almost identical. For large
$N$ the decays behave differently, with the brown and purple curves exhibiting nearly
uncorrelated noise behavior, and the green curve exhibiting more quasistatic
noise behavior.
Free evolution
decay for the brown and purple PSDs are different---exponential and Gaussian,
respectively, but their benchmarking decays are nearly identical.
The green curve has
$1/f^2$ PSD behavior at the pulsing frequency, and the red curve has $1/f$
PSD at the pulsing frequency; both
give rise to decays between uncorrelated and quasistatic.

Finally, we note that the relationship between RB sequence fidelity and Ising
model partition functions can be generalized to $d$-state systems.  The progression from
Eq.~(\ref{eq:singleQubitZ}) to Eq.~(\ref{eq:singleQubitPartitionFunctionSum})
for $d=2$ is structurally the same for general $d$: the non-trivial term
in the twirled free evolution matrix is the sum of cosines
over weights of the adjoint representation of \SUd;
products of cosine terms are converted into sums using the cosine addition
formula; the Hubbard-Stratonovich transformation converts the integral over
all error phases at different intervals into a partition function
sum over all possible configurations
of \SUd\ adjoint representation weights at $N$ sites. The details
of the analysis are given in the supplementary material.

In summary, we have shown that RB on one or more qubits has a deep
connection to the Ising model and admits long-range (power law) and
short range (exponential) decay behavior for noises with different spectra.  On
the surface, this implies that fitting RB experiments to an
exponential decay model can be dangerous and can lead to undependable
estimates of the error rate.  This can be mitigated by fitting only the
short sequence (linear decay) and asymptotic regime data,
which always gives a consistent estimate of
the average single gate error rate.

There is however a deeper issue.  For generic noise, the average gate error rate
does not uniquely determine the long time behavior.  Even if we can correctly
estimate the average gate error with RB we may
learn nothing about a system's
fault-tolerant behavior \cite{Kueng:2016jk}. Nonexponential
RB decay implies that there exist error correction/control
procedures that are more favorable than simply comparing the average gate error
rate to fault tolerance thresholds, thresholds that are generally calculated
using Markovian error models. A trivial example is that for perfect quasistatic
noise one could completely eliminate errors with decoupling sequences, but
whether there generically exist error mitigation techniques for correlated noise
is an interesting open problem.


\bibliography{Quantum_Computing}

\end{document}


\title{Randomized Benchmarking, Correlated Noise, and Ising Models:\\
Supplementary Material}
\author{Bryan H. Fong}
\email[]{bhfong@hrl.com}
\author{Seth T. Merkel}
\affiliation{HRL Laboratories, LLC, 3011 Malibu Canyon Road, Malibu, CA 90265}
\begin{abstract}
\end{abstract}
\maketitle
\section{High Effective Temperature Expansion for Partition Function}
Here we derive the approximate expression, Eq.~(11) of
the main text, for the
single qubit sequence fidelity partition function discussed in the main text.
Our starting point
is Eq.~(4) of the main text,
\begin{equation}
  \mathcal{Z} = \int_{\RN} d\bmt
  \left[
  \prod_{n=1}^{N}\frac{1}{3}\left(1+2\cos\theta_n\right)
  \right]
  \frac{e^{-\frac{1}{2}(\bmt-\bmm).\bmx^{-1}.(\bmt-\bmm)}}{\sqrt{(2\pi)^N |\bmx|}}.
\label{eq:singleQubitZ}
\end{equation}
We define new integration variables
$\bm{s} = (\bmt - \bmm)/\epsilon$, where $\epsilon\equiv\sqrt{2\beta}$
($\epsilon$ is the standard deviation of the random noise
in each free evolution interval). Here we are using $\beta$ to parameterize the
strength of the random noise $\bmx$ only; $\bmm$ can take finite values,
independent of the size of the random noise. With respect to the spin-one
Ising model,
$\beta$ only really represents an effective inverse temperature when
$\bmm$ also scales with $\beta$. By excluding $\beta$ scaling from $\bmm$, however,
we can obtain approximate
expressions that hold for small random noise (small $\beta$) and
finite mean $\bmm=\theta_0\allones$.

We expand the product in the square brackets in Eq.~(\ref{eq:singleQubitZ})
in an $\epsilon$ series,
\begin{eqnarray}
  \prod_{n=1}^N\frac{1+2\cos\theta_n}{3}
  &=& \left(\frac{1+2\cos\theta_0}{3}\right)^N
  \exp\left(-\epsilon\frac{2\sin\theta_0}{1+2\cos\theta_0}\allones.\bm{s}\right)
  \exp\left(-\epsilon^2\frac{2+\cos\theta_0}{(1+2\cos\theta_0)^2}\bm{s}.\bm{s}\right)
  \nonumber\\
  &&\times \left(1 -\epsilon^3 \frac{(7+2\cos\theta_0)\sin\theta_0}{3(1+2\cos\theta_0)^3}
  \sum_{n=1}^N s_n^3
  - \epsilon^4 \frac{28+12\cos\theta_0-12\cos2\theta_0-\cos3\theta_0}
  {12(1+2\cos\theta_0)^4}
  \sum_{n=1}^N s_n^4
  +\ldots
  \right).
\end{eqnarray}
This series expansion is obtained
by taking the exponential of the Taylor
series expansion of the logarithm of the left hand side product, keeping the first
and second orders in $\epsilon$ in the exponential, and re-expanding in
a Taylor series the exponential of all the remaining terms in the logarithm
expansion. Such an expansion plays a similar role to the
expansion of $e^{-\lambda \phi^4}$ in field theory.
The expansion
allows us to express all terms in the integral of Eq.~(\ref{eq:singleQubitZ})
as moments of a Gaussian
distribution with modified covariance matrix $\bmS$ and modified mean $\bmn$,
\begin{eqnarray}
  \bmS & \equiv & \bmx\left(\bmI + \frac{2(2+\cos\theta_0)}{(1+2\cos\theta_0)^2})
  \bmx\right)^{-1},\\
  \bm{y} & \equiv & 2\frac{\sin\theta_0}{1+2\cos\theta_0} \allones, \\
  \bmn & \equiv & -\bmS.\bm{y}.
\end{eqnarray}
Notice that $\bmS$ retains the $\beta$ scaling of $\bmx$ and that
$\bmn$ also is first order in $\beta$.
Performing the Gaussian moment integrations (Wick contractions)
and setting $\epsilon=1$
results in the following series expansion for the partition function:
\begin{eqnarray}
  \mathcal{Z} &=& \left(\frac{1+2\cos\theta_0}{3}\right)^N
  \frac{e^{\frac{1}{2}\bm{y}.\bmS.\bm{y}}}
  {\sqrt{\left|\bmI + \frac{2(2+\cos\theta_0)}{(1+2\cos\theta_0)^2}\bmx
  \right|}}
 \left(1 -\frac{(7+2\cos\theta_0)\sin\theta_0}{3(1+2\cos\theta_0)^3}
  \sum_{n=1}^N \left(3\Sigma_{nn}\nu_n+\nu_n^3\right)\right.
\nonumber\\&&
\quad \left.  - \frac{28+12\cos\theta_0-12\cos2\theta_0-\cos3\theta_0}
  {12(1+2\cos\theta_0)^4}
  \sum_{n=1}^N \left(3\Sigma_{nn}^2+6\Sigma_{nn}\nu_n^2+\nu_n^4\right)
  +\ldots  \right).
\label{eq:singleQubitZApproximate}
\end{eqnarray}

In the limit of zero mean noise $\theta_0=0$,
$\bmS=\bmx(I+\frac{2}{3}\bmx)^{-1}$ and $\bm{y}=\bmn=0$. The approximate
form for the partition function then becomes
\begin{equation}
  \mathcal{Z}_0 = \frac{1}{\sqrt{\left|\bmI + \frac{2}{3}\bmx\right|}}
  \left(1-\frac{1}{12}\sum_{n=1}^N\Sigma_{nn}^2+\ldots\right),
\label{eq:singleQubitZ0Approximate}
\end{equation}
Eq.~(11) in the main text.

\section{Fitting Quasistatic Noise Sequence Fidelity Decay to an Exponential}
In this section we illustrate the inconsistencies in fitting quasistatic sequence
fidelity decay to an exponential. We assume that the quasistatic
sequence fidelity decay is
described by the lowest order term in the approximate partition function,
Eq.~(13) of the main text,
with the nominal average gate error rate $\varepsilon = \frac{1}{3}\beta$.
With ``data'' generated from  Eq.~(13) of the main text, we
perform a weighted least-squares fit to an exponential functional form
minimizing the objective
\begin{equation}
\sum_{N=1}^{N_\mathrm{max}} \left(
\frac{1}{2}+\frac{1}{2}\frac{1}{\sqrt{1+4N \varepsilon}}
-A - B (1-2 K\varepsilon)^N
\right)^2 w_N,
\label{eq:fittingObjective}
\end{equation}
where $A$, $B$, and $K$ are the possible fitting parameters, and $N_\mathrm{max}$
is the maximum RB sequence length used in the fit. $K$ measures how far
the fit deviates from the nominal average gate error rate $\varepsilon$.

We consider four fitting scenarios:
\begin{enumerate}
\item $A$ and $B$ are fixed at $\frac{1}{2}$ and $w_N = 1$
\item $A$ and $B$ are fixed at $\frac{1}{2}$  and $w_N = \frac{1}{N}$
\item $A$ and $B$ are fitting parameters and $w_N=1$
\item $A$ and $B$ are fitting parameters and $w_N=\frac{1}{N}$.
\end{enumerate}
Scenarios 1 and 3 are equally weighted with respect to sequence length,
while scenarios 2 and 4 have shorter sequences more heavily weighted
than longer sequences, corresponding to fitting to sequence lengths that are
evenly sampled in $\log N$. For the figures that follow, we take $\beta=0.01$
and correspondingly, $1/(2\varepsilon)=150$. For exponential decay, $1/(2\varepsilon)$
corresponds to the sequence length giving the $1/e$-point.

\begin{figure}
\includegraphics[width=4in]{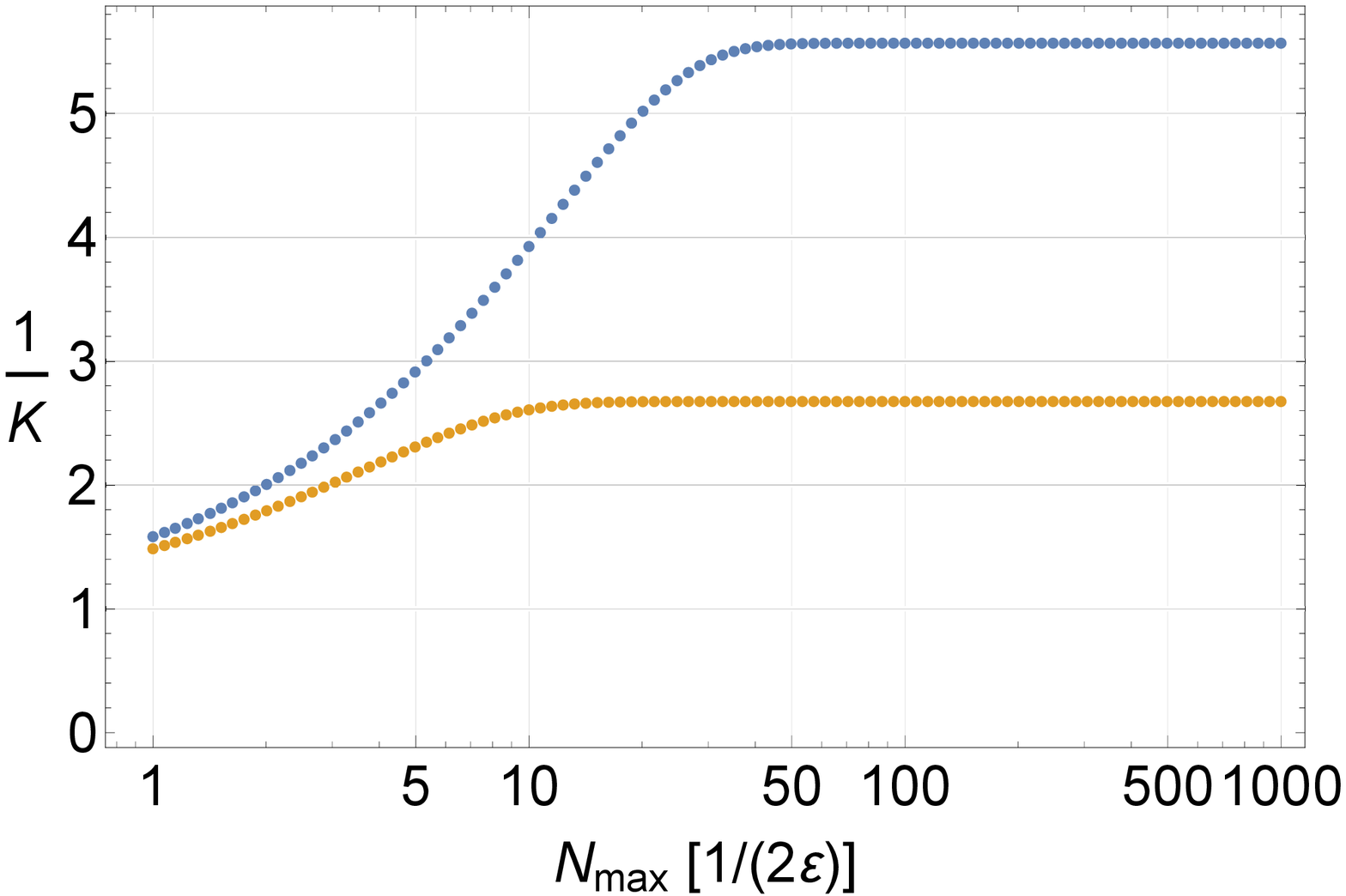}
\caption{\label{fig:fixedABUnderestimatePlot}
Plot of inverse fitting parameter $1/K$ versus maximum RB sequence length
$N_\mathrm{max}$
in units of inverse error rate $1/(2\varepsilon)=150$. Blue dots are $1/K$ for
fitting scenario 1 ($A=B=1/2$, $w_N=1$); orange dots are for scenario 2
($A=B=1/2$, $w_N=1/N$). For fixed $A$ and $B$, the average gate error rate
is always underestimated, by factors between $1.5$ and $5.6$, for maximum
sequence lengths between $1/(2\varepsilon)$ and $1000/(2\varepsilon)$.
}
\end{figure}

\begin{figure}
\includegraphics[width=4in]{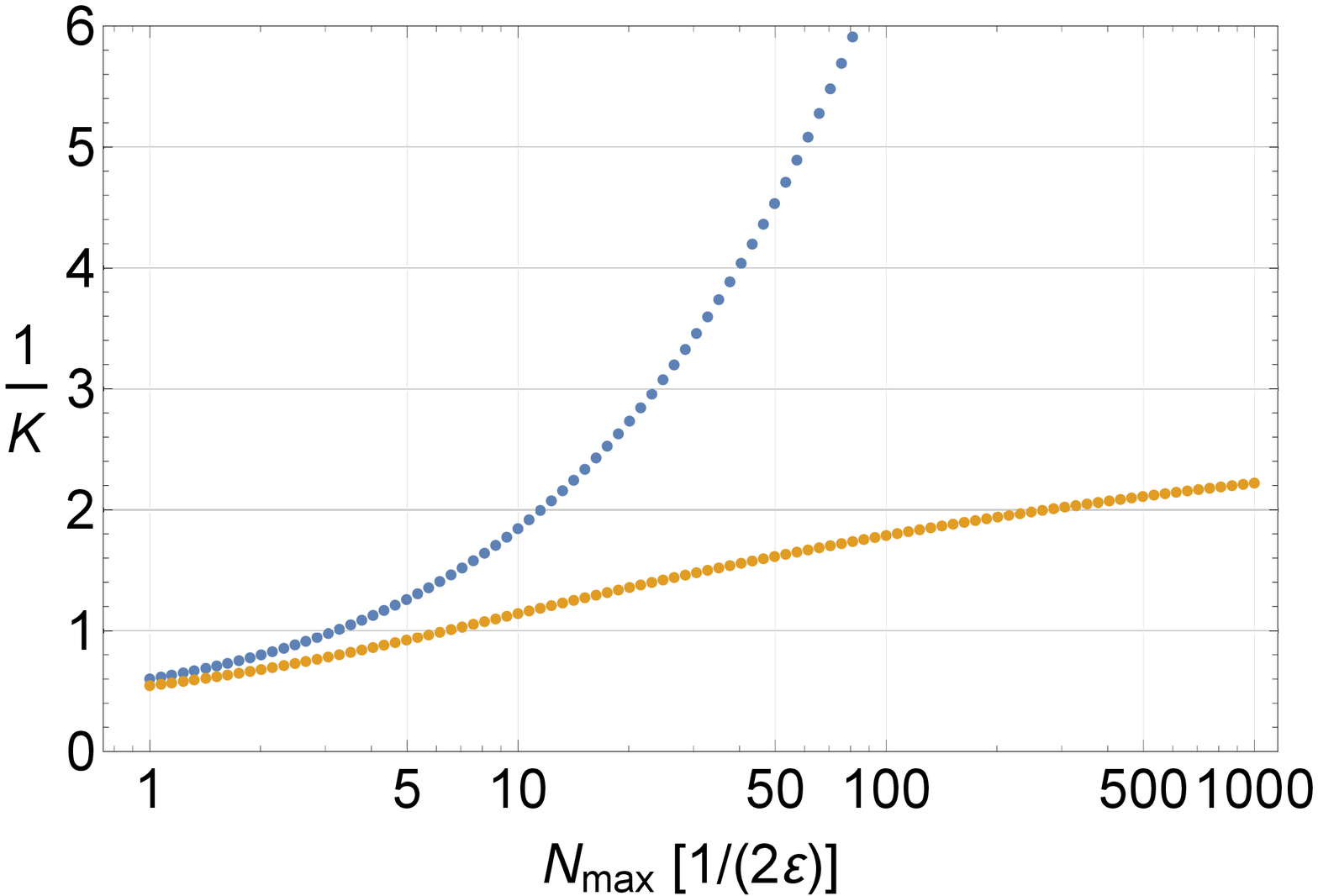}
\caption{\label{fig:floatingABUnderestimatePlot}
Plot of inverse fitting parameter $1/K$ versus maximum RB sequence length
$N_\mathrm{max}$
in units of inverse error rate $1/(2\varepsilon)=150$. Blue dots are $1/K$ for
fitting scenario 3 ($A$ and $B$ are fitting parameters, $w_N=1$);
orange dots are for scenario 4 ($A$ and $B$ are fitting parameters, $w_N=1/N$).
When $A$ and $B$ are allowed to vary, the average gate error rate is
overestimated for short maximum sequence lengths
$N_\mathrm{max}\sim1/(2\varepsilon)$ and underestimated for longer $N_\mathrm{max}$.
}
\end{figure}

Figures \ref{fig:fixedABUnderestimatePlot} and \ref{fig:floatingABUnderestimatePlot}
show the inverse of the error rate deviation parameter $K$ for the four fitting scenarios,
as a function of maximum RB sequence length $N_\mathrm{max}$ used in the fits.
Both over- and underestimation of the average gate error rate is possible,
dependent on the maximum sequence lengths used for fitting, and whether $A$ and
$B$ are allowed to vary. For scenarios 3 and 4, where $A$ and $B$ are allowed
to vary, $A$ and $B$ can differ substantially from their nominal values of $\frac{1}{2}$.
For scenario 3, equally weighted data, as $N_\mathrm{max}$ increases, the objective
is minimized by correctly fitting the nominal value of $A$, and severely
underestimating
both $B$ and the average gate error rate since most of the fitting data is at
large RB sequence lengths. For scenario 4, which weights shorter sequences
more heavily, $A$ and $B$ are approaching their nominal values as $N_\mathrm{max}$
increases, with the average gate error rate underestimated by a factor of $2.2$.
Note that the objective Eq.~(\ref{eq:fittingObjective}) is almost scale invariant for small
$\varepsilon$---the summand depends on the product
$N\varepsilon$ rather than $N$ and $\varepsilon$ separately. Figures 1--4
are essentially unchanged for different values of (small) $\varepsilon$.

Figure~\ref{fig:floatingABLogFitPlot} shows fits to the quasistatic sequence
fidelity for scenario 4, for four different values of $N_\mathrm{max}=
\{1,10,100,1000\}\times 1/(2\varepsilon)$. For $N_\mathrm{max}=1/(2\varepsilon)$
(blue curve in figure) the fit appears to be very good, but is incorrect:
$A=0.75$, $B=0.24$, and
$K=1.8$, giving a factor of $1.8$ overestimate of the average gate error rate.
Because $A$ and $B$ are allowed to vary, the initial nonexponential decay
can be fit by changing $A$ and $B$ from their nominal values.
When longer sequences are included in the
fitting data, the exponential functional fitting form no longer can be made to match
the full nonexponential decay.

\begin{figure}
\includegraphics[width=4in]{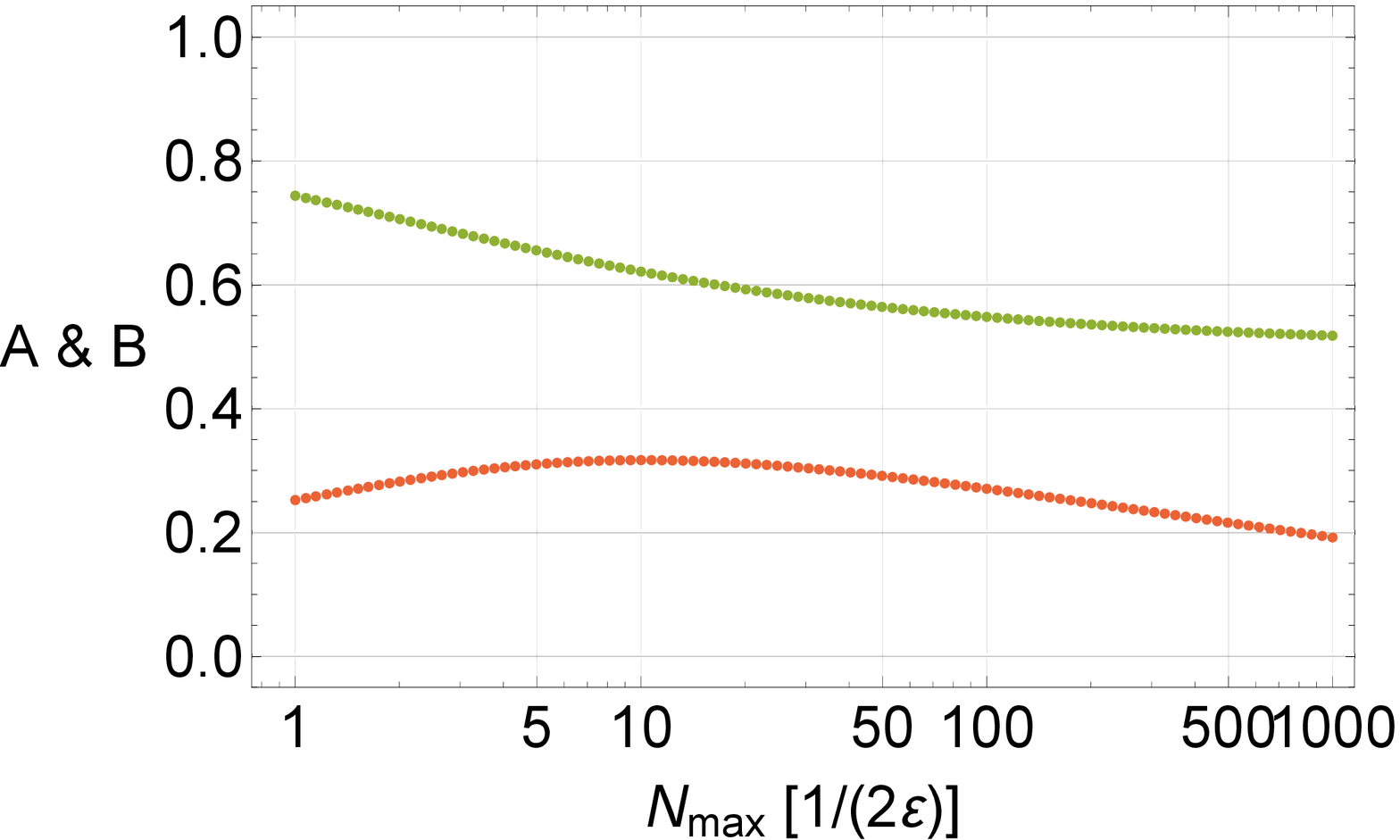}
\caption{\label{fig:floatingABLinearPlot}
Fitting parameters $A$ and $B$ versus maximum RB sequence length
$N_\mathrm{max}$
in units of inverse error rate $1/(2\varepsilon)=150$,
for fitting scenario 3 ($A$ and $B$ are allowed
to vary; $w_N=1$). Green dots are $A$ and red dots are $B$.
}
\end{figure}

\begin{figure}
\includegraphics[width=4in]{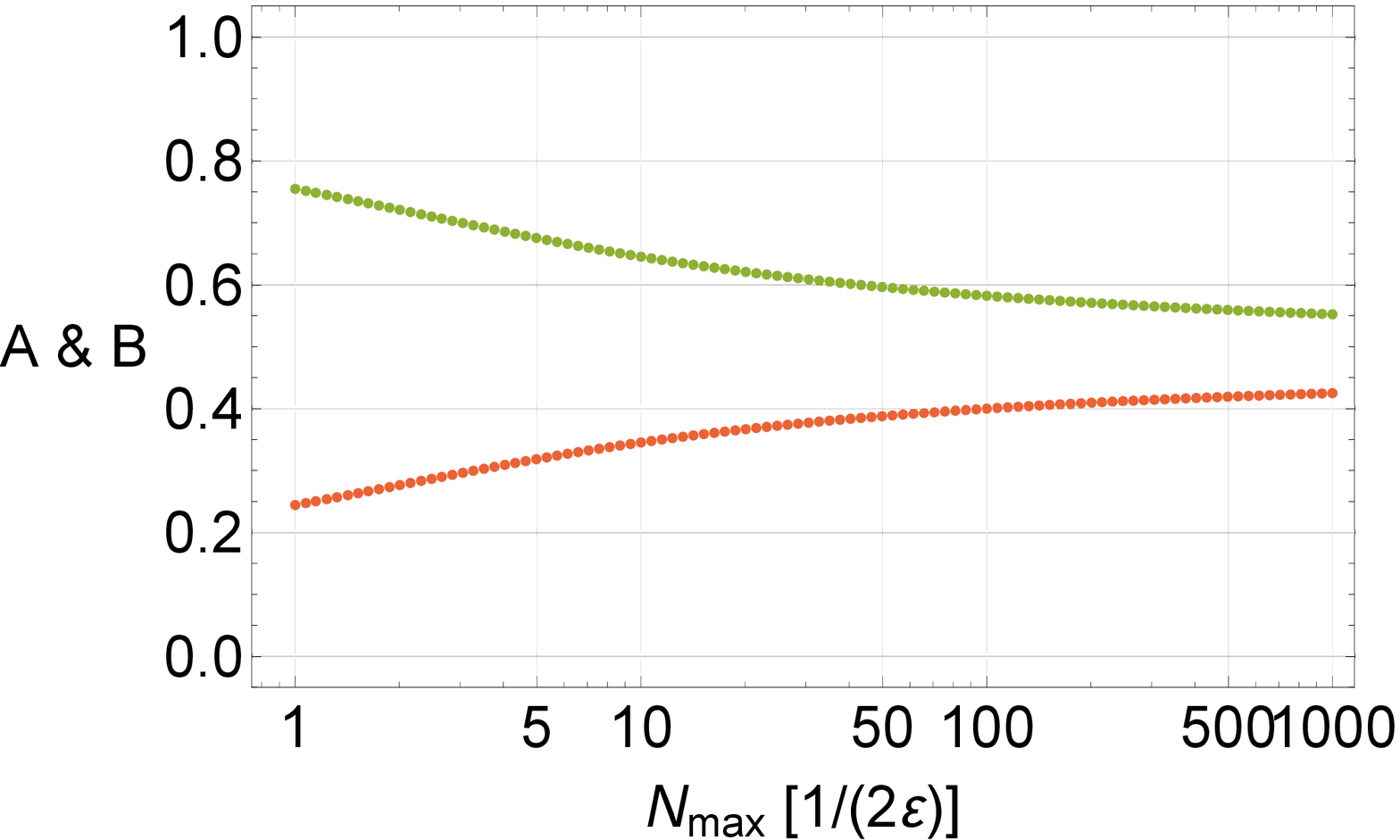}
\caption{\label{fig:floatingABLogPlot}
Fitting parameters $A$ and $B$ versus maximum RB sequence length
$N_\mathrm{max}$
in units of inverse error rate $1/(2\varepsilon)=150$,
for fitting scenario 4 ($A$ and $B$ are allowed
to vary; $w_N=1/N$). Green dots are $A$ and red dots are $B$.
}
\end{figure}

\begin{figure}
\includegraphics[width=4in]{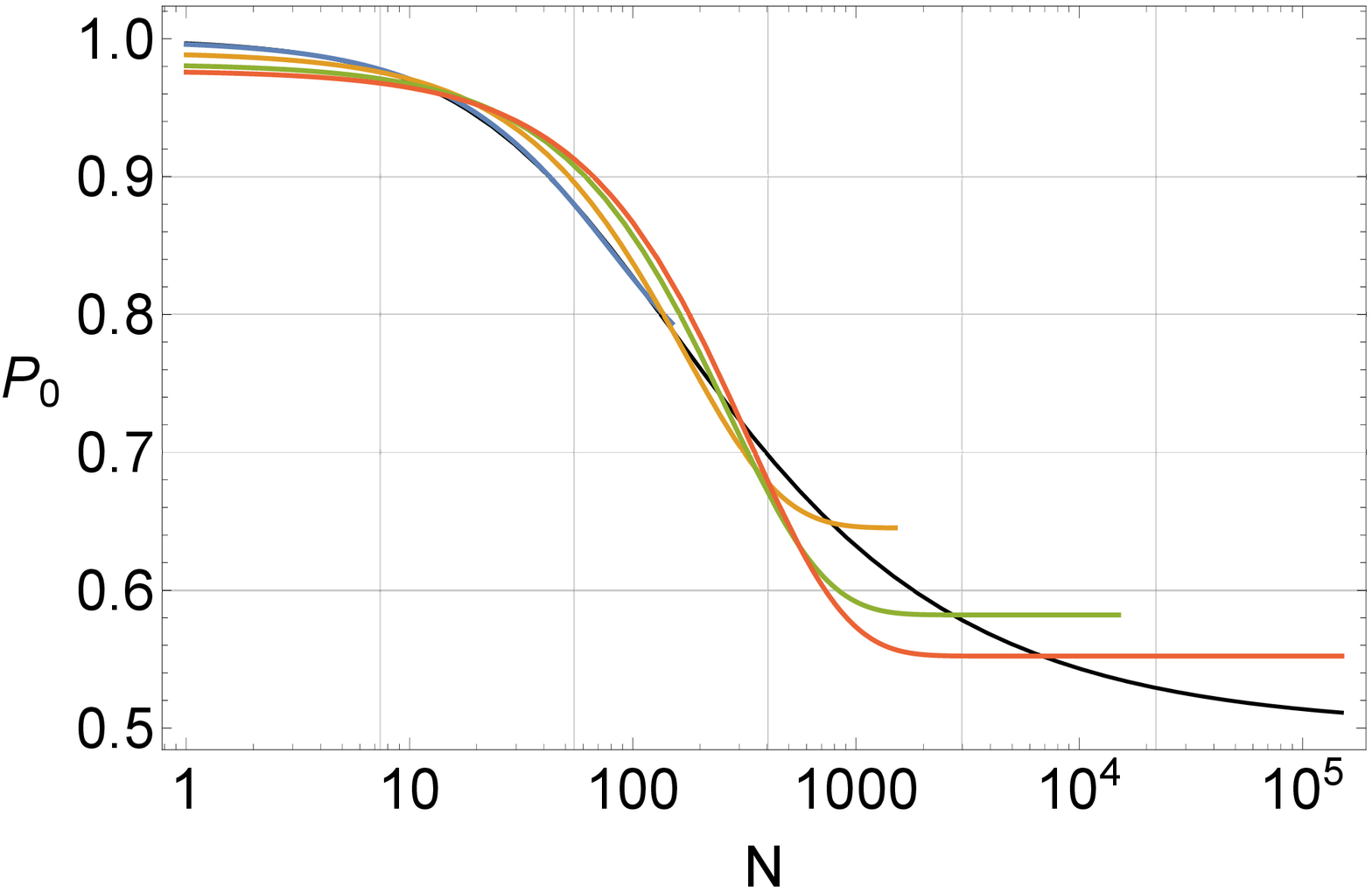}
\caption{\label{fig:floatingABLogFitPlot}
Sequence fidelity $P_0$ versus sequence length $N$, for
quasistatic ``data'' and fits from scenario 4 ($A$ and $B$ are
fitting parameters, $w_N=1/N$). Black curve is
approximate quasistatic sequence fidelity given by Eq.~(13) of the main text.
$\beta=0.01$ and $1/(2\varepsilon)=150$.
Blue fit has $N_\mathrm{max}=1/(2\varepsilon)$,
orange fit has $N_\mathrm{max}=10/(2\varepsilon)$,
green fit has $N_\mathrm{max}=100/(2\varepsilon)$,
and red fit has $N_\mathrm{max}=1000/(2\varepsilon)$. The fitting curves end at their
respective values of $N_\mathrm{max}$. The blue fit lies on top of the
black curve, but gives incorrect values for $A$, $B$, and $K$.
}
\end{figure}

\section{Benchmarking of $d$-State Systems}
Here we show that the analysis of single qubit benchmarking
and its relationship to the Ising model
can be extended to the benchmarking of $d$-state systems.
Following the derivation in the main text, we compute the expected benchmarking
sequence fidelity in Eq.~(1) of the main text, via the repeated application of
the twirled free evolution operator.
Because the Haar measure integration in the definition of the twirled free
evolution operator (Eq.~(2) of the main text) is translation invariant,
without loss of generality
we assume a diagonal free evolution unitary in $\Ud$,
\begin{equation}
F_n = \mathrm{diag}\left(e^{-i \theta_1^n},e^{-i \theta_2^n},\ldots,e^{-i \theta_d^n}\right),
\end{equation}
where $\theta_i^n$ is the error phase accumulated on the
$i^\mathrm{th}$ state in the $n^\mathrm{th}$ interval. (Note that the error
phase angle convention here differs from the standard SU(2) expression by
a factor of 2.) This free evolution
matrix
can be substituted into Eq.~(2) of the main text,
and the integration over the Haar measure
performed explicitly; however, a few group theoretical observations obviate the
need for explicitly performing $(d^2-1)$-dimensional integrals.

The action of the
twirled free evolution map on a density matrix is the product of the defining
and conjugate representations of \SUd, which in turn is the direct sum of the
(irreducible) trivial and adjoint representations. The twirled free evolution
map in any irreducible representation (irrep) commutes
with all elements in the irrep, again because of Haar measure invariance.
By Schur's lemma, the
twirled free evolution map must be proportional to the identity on each
irrep. The proportionality constant $K^{(n,\gamma)}$
for each irrep $D^{(\gamma)}$ is given by
\begin{equation}
\int_{U(d)}dU D^{(\gamma)}\left(U F_n U^{-1}\right)=K^{(n,\gamma)} D^{(\gamma)}(I).
\end{equation}
Taking the trace of both sides gives
$
K^{(n,\gamma)} = \chi^{(\gamma)}(F_n)/{d^{(\gamma)}},
$
where $\chi^{(\gamma)}(F_n)$ is the character of the free evolution operator
on the irrep $\gamma$, and $d^{(\gamma)}$ is the dimension of irrep $\gamma$.
The character $\chi^{(\gamma)}(F_n)$ can be written in terms of the weights
$w$ of irrep $\gamma$ and the angles $\theta_i^n$ of $F_n$ \cite{Elliott:1979,
Cahn:1984vn}. (The Weyl character formula is customarily used to express
$\chi^{(\gamma)}(F_n)$ in terms of the highest weight of $\gamma$, the Cartan
subalgebra, and a sum over the Weyl group, but to perform the integrations over
Gaussian distributed $\theta_i^n$ variables, it is more convenient to
express the character explicitly as a sum over all the weights.)

For the trivial and
adjoint irreps, in a basis
of $d^2$ $d\times d$ Hermitian matrices
$\{\lambda_i\}_{i=0}^{d^2-1}$ orthogonal with respect to the trace inner product,
 the matrix for the twirled free evolution map
is then
\begin{equation}
  R_{ij}^{(n)} =
\begin{cases}
1& \mathrm{for}\ i=j=0,\\
\frac{1}{d^2-1}\sum_{w\in W(d^2-1)}
   \cos\left(\sum_{k=1}^d w_k\theta_k^n\right)
&\mathrm{for}\ i=j=\{1,2,\ldots,d^2-1\},\\
0& \mathrm{otherwise},
\end{cases}
\end{equation}
where we assume that $\lambda_0 = I_{d\times d}$ (i.e., the upper left $1\times1$
block of $R^{(n)}$ corresponds to the trivial irrep, and the lower right
$(d^2-1) \times (d^2-1)$ block of $R^{(n)}$ corresponds to the adjoint irrep), and
$W(d^2-1)$ is the multiset of weights of the \SUd\ adjoint representation,
in the standard basis.
Adjoint representation weights in the standard basis are $d$-component vectors
with either all zero entries (of which there are $d-1$ repeated weights), or
two non-zero entries of $+1$ and $-1$ (of which there are $d(d-1)$ unique weights).

Repeated application of the twirled free evolution map to a pure state, again
assuming a Gaussian distribution for the noise, results in cosine-weighted
Gaussian integrals for the expected sequence fidelity. As in the $d=2$ case, we convert products
of cosines into sums using the cosine addition formula, and we again use the
(inverse) Hubbard-Stratonovich transformation to convert the
continuous $\theta_k^n$ error phase variables to discrete $w_i^m$ weight variables.
The final sequence fidelity expression is
\begin{widetext}
\begin{equation}
P_0 = \frac{1}{d}+\frac{d-1}{d}\frac{1}{(d^2-1)^N}
\sum_{w^1\in W}
\sum_{w^2\in W}\ldots
\sum_{w^N\in W}
\exp\left(-\frac{1}{2}\sum_{m,n=1}^N\sum_{i,j=1}^d
w_i^m\chi_{ij}^{mn}w_j^n + i \sum_{n=1}^N\sum_{i=1}^d w_i^n \theta_{0,i}^n\right).
\label{eq:generalP0}
\end{equation}
\end{widetext}
The tensor $\chi_{ij}^{mn}$ gives the covariance between the error phase at state $i$
and free evolution interval $m$,
and the error phase at state $j$ and free evolution interval $n$, and
$\theta_{0,i}^n$ is the mean error accumulated in state $i$ in interval $n$.
We again have an $N$ site Ising model partition function, where each site
variable is a $d$-dimensional vector, whose possible states are the weights
of the adjoint representation of \SUd. Specializing to $d=2$ recovers the
single qubit partition function of Eq.~(10) of the main text.

Approximate expressions for the sequence fidelity for $d$-state randomized
benchmarking can similarly be obtained through a high effective temperature
expansion of Eq.~(\ref{eq:generalP0}),
giving a lowest order formula in terms of $\chi_{ij}^{mn}$
analogous to the determinant expression in Eq.~(\ref{eq:singleQubitZ0Approximate}).

\bibliography{Quantum_Computing}